\documentclass[sigconf,nonacm]{acmart}
\usepackage{xspace}
\usepackage{balance}
\usepackage{tikz}
\usetikzlibrary{arrows.meta,positioning,fit,backgrounds,calc}
\usepackage{array}
\usepackage{booktabs}
\usepackage{subcaption}
\newcommand{\viol}{\textsf{violation}\xspace}
\newcommand{\noviol}{\textsf{no\_violation}\xspace}
\newcommand{\unk}{\textsf{unknown}\xspace}
\newcommand{\snd}{\textsf{intact}\xspace}
\newcommand{\sfac}{\textsc{sfac}\xspace}
\newcolumntype{R}[1]{>{\raggedright\arraybackslash}p{#1}}

\begin{document}

\title{Sound Distributed Runtime Verification under Incomplete Evidence}
\thanks{Accepted at the 28th International Conference on Distributed Computing and Networking (ICDCN 2027), Surathkal, India, January 4--7, 2027. This is the authors' version, released under CC-BY 4.0.}
\renewcommand{\shorttitle}{Sound Distributed Runtime Verification}

\author{Nikolaos Kekatos}
\orcid{0009-0000-2358-4270}
\affiliation{\institution{Clone Systems}\city{Larnaca}\country{Cyprus}}
\email{nkekatos@clone-systems.com}

\author{Antonio Anastasio Bruto da Costa}
\orcid{0000-0002-4590-0665}
\affiliation{\institution{Pure Silicon Solutions}\city{Birmingham}\country{United Kingdom}}
\email{antoniobdacosta@gmail.com}

\author{Alexios Lekidis}
\orcid{0000-0002-8243-7075}
\affiliation{\institution{University of Thessaly}\city{Larissa}\country{Greece}}
\email{alekidis@uth.gr}

\author{Pabitra Mitra}
\orcid{0000-0002-1908-9813}
\affiliation{\institution{Indian Institute of Technology Kharagpur}\city{Kharagpur}\country{India}}
\email{pabitra@cse.iitkgp.ac.in}

\author{Hosein Nazarpour}
\orcid{0009-0000-4614-6003}
\affiliation{\institution{Amazon}\city{Paris}\country{France}}
\email{nazarpur@gmail.com}

\author{Priyanka Sinha}
\orcid{0000-0002-7971-4449}
\affiliation{\institution{Docyt}\city{Kolkata}\country{India}}
\email{priyanka.sinha.iitg@gmail.com}

\renewcommand{\shortauthors}{Kekatos et al.}

\begin{abstract}
Distributed runtime verification often requires combining evidence from several nodes to check properties that cannot be evaluated locally. However, when messages are delayed, lost, or deliberately suppressed, the verifier may not know whether a violation occurred or whether the relevant evidence is missing. This creates a risk of incorrectly reporting compliance. We study the requirements for sound monitoring under these conditions. We first show that a deadline-bounded monitor cannot be simultaneously sound, decisive, and available when an adversary can suppress evidence. The monitor must therefore be able to abstain. We then distinguish two sources of incomplete evidence. Sequence numbers and periodic reports can identify missing events, whereas detecting silent streams requires timing assumptions. We also examine how membership changes and coordinator failures affect verdict soundness and recovery. We implement these mechanisms in a three-tier monitoring architecture using MQTT and NATS JetStream. Experiments with an LLM-planned robot swarm demonstrate the failure modes, show how the proposed mechanisms prevent unsupported compliance verdicts, and measure their runtime and communication costs.
\end{abstract}



\maketitle

\section{Introduction}
\label{sec:intro}

Runtime verification checks a running system against formal specifications, using the events the
system emits during execution~\cite{leucker2009brief}. When the system is distributed, the events that
decide a requirement come from several nodes, and no single node sees enough of them to reach a verdict
on its own~\cite{bauer2012decentralised,colombo2016organising,francalanza2018distributed,sen2004decentralized}.
A data platform that must enforce a retention policy across services is one example. A fleet of
autonomous vehicles that must respect a limit on the fleet as a whole is
another~\cite{chung2018aerial,ferrando2021rosmonitoring,garg2011audit}.

The usual answer is to forward the events of every node to a central verifier and evaluate the global
specification there. The verdict is then only as good as the evidence that reached the verifier.
Messages are delayed, lost, or suppressed on purpose, and from the verifier's side an idle node and a
node whose reports never arrived look the same~\cite{basin2015failure,basin2017outoforder,joshi2017lossy}.
In security monitoring this gap matters most, because the events an adversary wants to hide are exactly
the ones that would reveal a violation.

In this work, we consider an adversary who can interfere with the delivery of evidence but cannot forge
or alter the events the monitored nodes produce. Evidence that does arrive can therefore prove a violation. Proving
compliance is harder. A received trace without a violation says nothing about compliance while relevant
events may still be missing, so a monitor needs to know whether its evidence is complete before it
commits to a definite verdict.

Earlier work covers decentralised monitoring, monitoring under process crashes, and the evaluation of
incomplete or out-of-order
streams~\cite{basin2015failure,basin2017outoforder,bonakdarpour2016decentralized,elhokayem2017decentralized,fraigniaud2014opinions,joshi2017lossy}.
Failure-detector theory describes what processes in an asynchronous system can learn about one
another~\cite{chandra1996unreliable,chandra1996weakest}. We ask what these problems demand of the
evidence-delivery mechanism itself, that is, which failures can be detected at all, when the evidence at
hand supports a sound verdict, and what else nodes and verifier must exchange.
Fig.~\ref{fig:pipeline} shows the evidence pipeline and the failure modes we consider.

The paper makes three contributions.
\begin{itemize}
\item \textbf{Limits of sound monitoring.} For the class of cross-node properties we consider, a
  deadline-bounded monitor cannot be sound, decisive, and available at the same time once an adversary
  can suppress evidence (Theorem~\ref{thm:trilemma}). A monitor that stays sound must be allowed to
  return an inconclusive verdict (Corollary~\ref{cor:three}).
\item \textbf{Evidence completeness and failure detection.} Missing events and silent streams are
  different failures and call for different detectors (Lemma~\ref{lem:loss} and
  Theorem~\ref{thm:detector}). We give the reporting cost of accountable negative verdicts
  (Proposition~\ref{prop:cost}) and show that membership changes have to travel as evidence
  (Theorem~\ref{thm:membership}).
\item \textbf{Protocol design and recovery.} A three-tier monitoring protocol implements these
  mechanisms, and we prove it sound (Proposition~\ref{prop:protocol}). We then destroy the coordinator
  and work out how sequence numbering, verdict stamping, and acknowledgements decide what survives
  (Theorem~\ref{thm:attrition}).
\end{itemize}
The evaluation runs the monitors of an LLM-planned surveillance swarm over MQTT and NATS JetStream.
The experiments cover missing evidence, node silencing, membership changes, coordinator destruction,
and reporting bandwidth. Over the fault campaign the protocol produced no silent false all-clear, where
a best-effort coordinator produced three. Measurements on the live brokers and on synthetic traces give
runtime cost and scalability.

Section~\ref{sec:model} fixes the system model and the monitoring properties.
Sections~\ref{sec:impossible} and~\ref{sec:detectors} contain the theoretical requirements,
Section~\ref{sec:protocol} the protocol, and Section~\ref{sec:extensions} the treatment of membership
changes and coordinator destruction. Section~\ref{sec:eval} reports the experiments,
Section~\ref{sec:related} reviews related work, and Section~\ref{sec:conclusion} concludes.

\newlength{\bxw}\newlength{\bxtmp}
\setlength{\bxw}{27mm}
\newcommand{\bxmax}[1]{\settowidth{\bxtmp}{#1}\ifdim\bxtmp>\bxw\setlength{\bxw}{\bxtmp}\fi}
\bxmax{\small node $a_i$}       \bxmax{\scriptsize local monitor}
\bxmax{\small coordinator $c$}  \bxmax{\small verifier $v$}
\bxmax{\small verdict $(s,c)$}  \bxmax{\scriptsize + provenance}

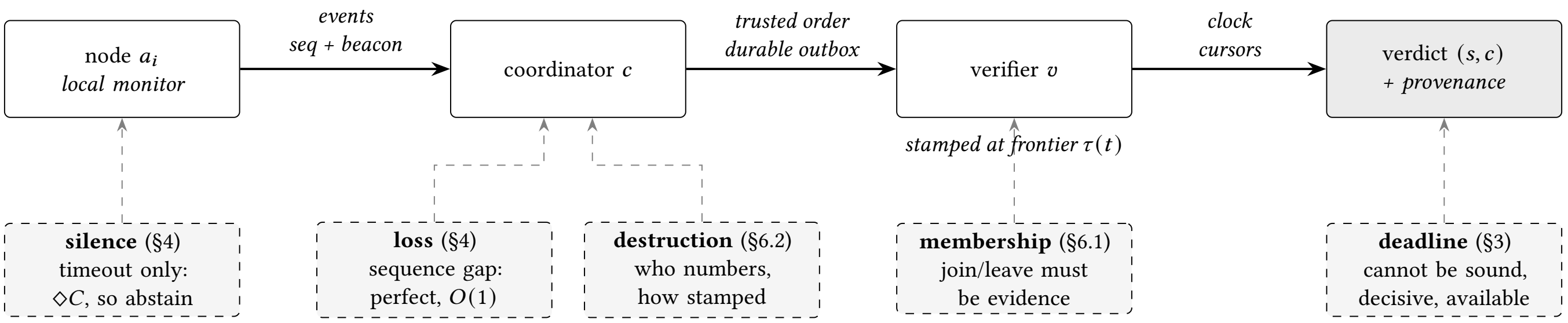
\begin{figure*}[t]
\centering
\resizebox{\textwidth}{!}{%
\begin{tikzpicture}[font=\small,>=Stealth,
  box/.style={draw,rounded corners=2pt,align=center,minimum height=12mm,text width=\bxw,
              inner sep=3pt},
  mech/.style={font=\small\itshape,align=center},
  fault/.style={draw,dashed,rounded corners=2pt,align=center,font=\small,
                minimum height=10mm,text width=27mm,inner sep=3pt,fill=black!4},
  a/.style={-Stealth,thick},
  f/.style={-Stealth,dashed,gray}]

  \node[box](node){node $a_i$\\{\small\itshape local monitor}};
  \node[box,right=26mm of node](coord){coordinator $c$};
  \node[box,right=26mm of coord](ver){verifier $v$};
  \node[box,right=24mm of ver,fill=black!8](verdict){verdict $(s,c)$\\{\small\itshape + provenance}};

  \draw[a](node)--node[mech,above=1pt]{events\\seq + beacon}(coord);
  \draw[a](coord)--node[mech,above=1pt]{trusted order\\durable outbox}(ver);
  \draw[a](ver)--node[mech,above=1pt]{clock\\cursors}(verdict);

  \node[mech,below=1mm of ver]{stamped at frontier $\tau(t)$};

  \node[fault,below=13mm of node](f1){\textbf{silence} (\S\ref{sec:detectors})\\
        timeout only: $\Diamond C$, so abstain};
  \node[fault,below=13mm of ver](f4){\textbf{membership} (\S\ref{sec:membership})\\
        join/leave must be evidence};
  \node[fault,below=13mm of verdict](f5){\textbf{deadline} (\S\ref{sec:impossible})\\
        cannot be sound,\\decisive, available};
  \coordinate (lx) at ($(node)!0.70!(coord)$);
  \coordinate (dx) at ($(coord)!0.30!(ver)$);
  \node[fault] at (lx |- f1)(f2){\textbf{loss} (\S\ref{sec:detectors})\\
        sequence gap: perfect, $O(1)$};
  \node[fault] at (dx |- f1)(f3){\textbf{destruction} (\S\ref{sec:attrition})\\
        who numbers, how stamped};

  \draw[f](f1)--(node); \draw[f](f4)--(ver); \draw[f](f5)--(verdict);
  \draw[f](f2.north)--++(0,7mm)-|([xshift=-3mm]coord.south);
  \draw[f](f3.north)--++(0,7mm)-|([xshift=3mm]coord.south);

\end{tikzpicture}}
\caption{The evidence pipeline, and where each result applies. Evidence flows left to right, and the
dashed boxes are the four ways it can fail to arrive, or to mean what the verifier takes it to mean.
The deadline on the right is the obstruction that makes abstention necessary.}
\label{fig:pipeline}
\end{figure*}

\section{System Model and Problem Definition}
\label{sec:model}

\paragraph{Nodes, streams and the verifier.} A system is a set of nodes $A=\{a_1,\dots,a_n\}$ and a
verifier $v$. Each node observes local activity and emits \emph{events}. An event is a tuple
$e=(\mathit{sq},\tau,a,k,\mathit{attrs})$ where $\mathit{sq}$ is a per-node monotone sequence
number, $\tau$ a timestamp that does not decrease along a node's sequence, $a\in A$ the emitting node, $k$ an event kind and $\mathit{attrs}$ typed
attributes. Cross-node order is fixed by the emission timestamps $\tau$ with $(a,\mathit{sq})$ breaking ties,
so it is defined on every event of $T$, including those the adversary removes before they reach a
coordinator. The trusted ingest order of Sect.~\ref{sec:protocol} is the verifier's observable proxy
for it, and a disagreement between the two is detectable rather than silently absorbed. Chains whose
steps fall closer together than the clock-skew bound are outside the class.
The local stream of $a$ is the sequence of events it emits and the \emph{true trace} $T$
is their union. The verifier consumes a \emph{received} stream $R\subseteq T$, and monitoring
proceeds in \emph{ticks}: at each tick the monitor must produce, for each property under
observation, a verdict from the domain defined below.

\paragraph{Communication and adversary.} Message passing is asynchronous: messages may be delayed
arbitrarily but not forever unless dropped. The adversary may drop, delay and reorder messages on
any link, and may \emph{silence} any node indefinitely by jamming it or partitioning it from the
verifier. An emitted event is \emph{missing} while the verifier does not hold it, and \emph{lost} if
it is never delivered; a node is \emph{silent} in a window if the verifier receives nothing from it. Nodes are authenticated and non-Byzantine in the core model: a node does not forge or
alter the events it emits, so $R\subseteq T$ always: the adversary can suppress or delay evidence but cannot modify event contents. Integrity faults, where a node under-reports its own
activity, need an independent corroborating witness and are treated in Sect.~\ref{sec:membership}.

\paragraph{The monotone-witness property class.} A property $P$ over traces is
\emph{monotone-witness} if there is a set $\mathcal{W}(P)$ of finite \emph{witness sets} such that
$P$ holds of a trace $T$ if and only if $W\subseteq T$ for some $W\in\mathcal{W}(P)$. Equivalently, a
finite set of events can establish $P$, and adding events cannot invalidate an existing witness.
Examples include \emph{existential conjunction} (several kinds of evidence concerning the same
object), \emph{threshold} (at least $m$ distinct nodes satisfying a predicate within a window),
\emph{sum} (an aggregate over nodes exceeding a budget), and \emph{ordered chain} (steps taken in a
prescribed order across nodes). Throughout the paper, $P$ denotes an incident condition, so the monitor
\emph{fires} when $P$ holds.

\paragraph{Absence claims.} Coverage obligations require additional evidence. The claim that a sector
was not observed is not monotone-witness over ordinary event traces: a later report of an observation
within the relevant window can invalidate it. We address this by treating the sequence reports of
Sect.~\ref{sec:detectors} as events. Commitments from all members, covering the window, can then
provide a finite witness that no observation occurred. A commitment binds a sequence high-water mark
to a time frontier (Sect.~\ref{sec:detectors}). Each commitment must be final for its node and
window, meaning that no subsequent event may appear within the committed sequence prefix. Under the
non-Byzantine assumption, a \emph{well-formed} extension cannot contain both these commitments and an
additional observation from the committed interval. Monotonicity therefore holds over well-formed
extensions that respect the commitments, rather than arbitrary supersets of events. Theorem~\ref{thm:trilemma}
is stated for monotone-witness properties; coverage obligations are covered only through this
construction, which also requires that the membership be complete for the window
(Sect.~\ref{sec:membership}).

\paragraph{Verdict domain.} A verdict is a pair $(s,c)$. The security axis is
$s\in\{\viol,\noviol,\unk\}$; the completeness axis is
$c\in\{\snd\succ\textsf{degraded}\succ\textsf{incomplete}\succ\textsf{unavailable}\}$, ordered by
how much of the evidence the verifier can account for. The
completeness of a composed verdict is the
\emph{meet} of the completeness of its contributing streams, and the \emph{downgrade rule} maps
$(\noviol,c)$ with $c\prec\snd$ to \unk: a negative finding is reported as such only when the
evidence supporting it is accounted for. Positive findings need no such treatment, since by
$R\subseteq T$ a witness that arrived is a witness that occurred. Emitting \unk we call
\emph{abstention}.

\paragraph{What a monitor must not do.} Let the \emph{oracle} verdict be the verdict computed on the
true trace $T$. The failure of interest is a false compliance verdict caused by suppressed evidence.

\begin{definition}[Silent false all-clear]
\label{def:sfac}
A monitor commits a \emph{silent false all-clear} (\sfac) at a tick if it outputs $(\noviol,\snd)$
while the oracle verdict at that tick is \viol. We also write \emph{silent all-clear} for this.
\end{definition}

An \sfac is worse than a missed detection, because to whoever consumes the verdict it is
indistinguishable from a genuinely clean system: it converts an attack into positive assurance. We
can now name the three properties Sect.~\ref{sec:impossible} shows are jointly unattainable.

\begin{definition}[Sound, decisive, available]
\label{def:sda}
A monitor is \emph{sound} if it never commits an \sfac and never reports \viol when the oracle
verdict is \noviol; \emph{decisive} if it never outputs \unk; and \emph{available} if it answers by
every tick's deadline. It is \emph{live} if, whenever no event is missing, it outputs the oracle
verdict within a bounded delay.
\end{definition}

Liveness excludes the trivial monitor that always outputs \unk, which is sound and available.

Soundness is two-sided: it forbids manufacturing an all-clear and manufacturing an alarm, so a
monitor cannot escape the \sfac by alarming on every gap.

\section{Limits of Sound Distributed Monitoring}
\label{sec:impossible}

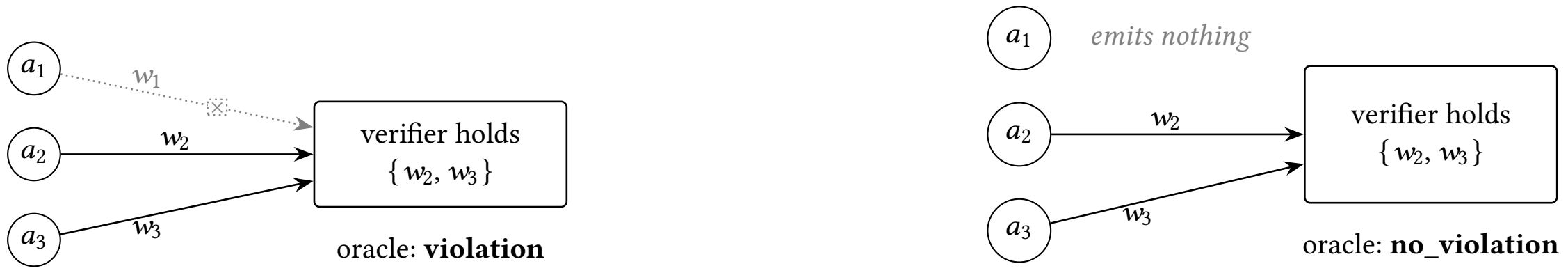
\begin{figure*}[t]
\centering
\begin{subfigure}[t]{0.47\textwidth}
\centering
\begin{tikzpicture}[font=\small,>=Stealth,
  n/.style={draw,circle,inner sep=0pt,minimum size=5mm,line width=.4pt},
  v/.style={draw,rounded corners=1.5pt,minimum height=10mm,minimum width=24mm,align=center,
            line width=.5pt},
  a/.style={-Stealth,line width=.5pt},
  dead/.style={-Stealth,line width=.5pt,densely dotted,black!50},
  vl/.style={font=\small}]
  \node[n](a1){$a_1$};
  \node[n,below=3mm of a1](a2){$a_2$};
  \node[n,below=3mm of a2](a3){$a_3$};
  \node[v,right=24mm of a2](av){verifier holds\\$\{w_2,w_3\}$};
  \draw[dead](a1)--node[pos=.34,above,inner sep=1pt]{$w_1$}
        node[pos=.62,draw,fill=white,inner sep=.6pt,line width=.4pt]{\tiny$\times$}(av);
  \draw[a](a2)--node[pos=.45,above,inner sep=1pt]{$w_2$}(av);
  \draw[a](a3)--node[pos=.34,below,inner sep=1pt]{$w_3$}(av);
  \node[vl,below=1.6mm of av]{oracle: \textbf{violation}};
\end{tikzpicture}
\caption{$E_1$: all three witnesses occur, and $w_1$ is jammed in transit.}
\label{fig:e1}
\end{subfigure}\hfill
\begin{subfigure}[t]{0.47\textwidth}
\centering
\begin{tikzpicture}[font=\small,>=Stealth,
  n/.style={draw,circle,inner sep=0pt,minimum size=6mm,line width=.4pt},
  v/.style={draw,rounded corners=1.5pt,minimum height=13mm,minimum width=24mm,align=center,
            line width=.5pt},
  a/.style={-Stealth,line width=.5pt},
  dead/.style={-Stealth,line width=.5pt,densely dotted,black!50},
  vl/.style={font=\small}]
  \node[n](b1){$a_1$};
  \node[n,below=3mm of b1](b2){$a_2$};
  \node[n,below=3mm of b2](b3){$a_3$};
  \node[v,right=24mm of b2](bv){verifier holds\\$\{w_2,w_3\}$};
  \node[right=2.5mm of b1,font=\small\itshape,black!50]{emits nothing};
  \draw[a](b2)--node[pos=.45,above,inner sep=1pt]{$w_2$}(bv);
  \draw[a](b3)--node[pos=.34,below,inner sep=1pt]{$w_3$}(bv);
  \node[vl,below=1.6mm of bv]{oracle: \textbf{no\_violation}};
\end{tikzpicture}
\caption{$E_2$: $w_1$ never occurs.}
\label{fig:e2}
\end{subfigure}
\caption{Two executions a verifier cannot distinguish. The verifier holds $\{w_2,w_3\}$ in both,
so a monitor obliged to answer at each tick errs in one of them (Thm.~\ref{thm:trilemma}), and
soundness requires the third verdict value.}
\label{fig:indist}
\end{figure*}

Under the model of Sect.~\ref{sec:model}, Fig.~\ref{fig:indist} shows two executions that produce
identical observations at the verifier but require different oracle verdicts, so no monitor can be
sound, decisive, and available at once.

\begin{theorem}[Trilemma]
\label{thm:trilemma}
Let $P$ be a monotone-witness property with a witness set requiring events from at least two nodes.
In the model of Sect.~\ref{sec:model}, no monitor for $P$ is simultaneously sound, decisive and
available.
\end{theorem}

\begin{proof}
Fix a $\subseteq$-minimal witness set $W\in\mathcal{W}(P)$, which exists since witness sets are
finite, and a node $a_i$ contributing an event $w\in W$. Consider two executions in which the nodes
emit the events of $W$ and nothing else. In $E_1$, every node emits its events of $W$, so $P$ holds
of the true trace and the oracle verdict is \viol; the adversary drops the message carrying $w$, or
equivalently silences $a_i$ immediately after it emits $w$. In $E_2$, the nodes behave as in $E_1$
except that $a_i$ emits nothing, so the true trace is $W\setminus\{w\}$. Any witness set contained
in it would be a proper subset of $W$, which minimality excludes, so the oracle verdict is \noviol;
no message carrying $w$ is sent, and $a_i$ is not silenced.

The verifier's local state is identical in $E_1$ and $E_2$ at every finite time. In both it has
received exactly the events of $W\setminus\{w\}$ and nothing from $a_i$, and asynchrony means
``$w$ has not arrived yet'' and ``$w$ will never arrive'' are not distinguishable from that state:
any message that would distinguish them is either not sent ($E_2$) or dropped ($E_1$).

Let $d$ be a tick deadline. Availability requires a verdict by $d$ and decisiveness forbids \unk, so
the output is \viol or \noviol, and because the local states coincide it is the same in both
executions. If it is \noviol then in $E_1$ it is $(\noviol,c)$ against an oracle verdict of \viol,
and soundness forces $c=\snd$, since the downgrade rule would otherwise have produced \unk,
contradicting decisiveness; that is an \sfac. If the output is \viol then in $E_2$ the monitor
reports a violation against an oracle verdict of \noviol, contradicting the second half of
soundness. Either way soundness fails.
\end{proof}

\begin{corollary}[Three values are necessary]
\label{cor:three}
Any sound and available monitor for such a $P$ has a verdict domain with at least three
distinguishable outputs, one of which is reached both in executions where the oracle verdict is
\viol and in executions where it is \noviol.
\end{corollary}

\paragraph{What escapes the theorem.} Three assumptions carry it. \emph{Locality:} if every witness
set lies within one node, no evidence need travel. \emph{Asynchrony:} under bounded delay, silence past
the bound is informative, the escape Sect.~\ref{sec:detectors} formalises as a detector.
\emph{Deadlines:} a monitor that need only be eventually correct can wait, which a mission deadline
forbids.

\paragraph{Relation to FLP and CAP} The result follows from neither. FLP concerns asynchronous consensus
with one faulty process~\cite{fischer1985impossibility}; here a single verifier decides, and the
obstruction is that the decision depends on evidence the adversary controls. CAP~\cite{gilbert2002brewer}
trades consistency against availability under partition; the trilemma trades \emph{soundness} against
decisiveness, with a tick deadline in the role of partition tolerance.

\section{Evidence Completeness and Failure Detection}
\label{sec:detectors}

Theorem~\ref{thm:trilemma} establishes that sound and available monitoring sometimes requires
abstention. We now examine when definite verdicts are possible and how the required detection
mechanisms differ between missing events and silent streams.

\begin{lemma}[Gaps certify missing evidence]
\label{lem:loss}
Let each node number its events contiguously, and let the verifier record, per stream, which numbers
it holds. If a stream delivers $j+2$ while $j+1$ is not held, then $j+1$ was emitted and is missing.
The verifier raises no false positive. With FIFO delivery per stream the state is the highest number
received, $O(1)$; with reordering it is a buffer of out-of-order numbers, $O(w)$ for a reordering
window $w$.
\end{lemma}

\begin{proof}
Numbers are contiguous, so $j+2$ exists only if $j+1$ was emitted. A gap can only be created by an
emitted event, and $R\subseteq T$ means the verifier never invents one.
\end{proof}

A gap shows that evidence is missing, not that it is lost: the event may still be in flight, and under
unbounded delay the two cannot be separated. A loss in a stream's quiescent tail leaves no gap;
Proposition~\ref{prop:cost} closes that case.

Lemma~\ref{lem:loss} gives a \emph{perfect} detector for incompleteness in the sense of the
failure-detector literature~\cite{chandra1996unreliable}: it never suspects wrongly and eventually
suspects every gap. Proposition~\ref{prop:cost} closes the quiescent-tail case and prices
both detectors.

\begin{definition}[Evidence-completeness detector]
\label{def:C}
For a node $a$ and a window $w$, the detector $C$ reports whether $a$ emitted an event in $w$ that
the verifier does not hold. $C$ is \emph{perfect} if it never errs, and \emph{eventually perfect},
$\Diamond C$, if it errs only over a finite prefix.
\end{definition}

A deployment approximates $C$ with two mechanisms. Numbering and periodic sequence reports expose
missing events; timeouts over beacons expose unavailable streams under partial synchrony. $C$ is
strictly stronger than a crash detector~\cite{chandra1996unreliable}, which does not suffice, since
the adversary may drop a message on a live node's link.

\begin{theorem}[Silence needs an eventually-perfect detector]
\label{thm:detector}
In the asynchronous model, a silenced node and an idle compliant one cannot be told apart from the
event stream alone. For monotone-witness properties:
\emph{(i)} given $C$, a sound monitor exists that returns a definite verdict whenever $C$ reports no
missing event, and abstains only when it does;
\emph{(ii)} given only $\Diamond C$, a monitor bound by a tick deadline cannot be both sound and
decisive, so it must abstain; and
\emph{(iii)} a sound, decisive, deadline-bounded monitor for the single-stream property ``this node
emitted an event in the current window'' is a perfect $C$ for that node.
\end{theorem}

\begin{proof}
\emph{(i)} Let $R$ be the received stream. If $P$ holds on $R$, report \viol, which is sound since
$R\subseteq T$. Otherwise, if $C$ reports no missing event for any contributing node, $R$ and $T$
agree on those streams, so $P$ fails on $T$ and \noviol is correct. If $C$ reports a missing event,
it may complete a witness, and the monitor abstains. The detector need not say which event is
missing.
\emph{(ii)} At any finite time a $\Diamond C$ suspicion may lie in its unreliable prefix, so
attributing silence to a failure, or to idleness, may be wrong. Deciding anyway violates one half of
soundness, and refusing violates decisiveness.
\emph{(iii)} A sound decisive verdict on that property at each deadline is a correct report of whether
the node has unaccounted evidence. The converse concerns the single-stream case and does not place
$C$ in the detector hierarchy.
\end{proof}

\begin{table}[t]
\caption{Failure classes differ in detectability, completeness value and cost. The two mechanisms do not
substitute for one another (ablation, Sect.~\ref{sec:eval}).}
\label{tab:detect}
\small
\begin{tabular}{@{}R{1.45cm}R{1.85cm}R{1.6cm}R{1.6cm}@{}}
\toprule
\textbf{Failure} & \textbf{Detector} & \textbf{Verdict} & \textbf{Cost} \\
\midrule
event loss & perfect, sequence gap & \textsf{incomplete} & $O(1)$/stream \\
silence & $\Diamond C$, timeout & \textsf{unavailable} & $O(n)$ liveness state \\
reordering & perfect, trusted ingest order & \textsf{degraded} & $O(\text{buffer})$ \\
membership change & only if reported as evidence & \unk on thresholds & $O(1)$/change \\
\bottomrule
\end{tabular}
\end{table}

Table~\ref{tab:detect} summarises the picture. A composed verdict is only as accountable as its least
accountable contributing stream, so completeness is a lattice and composition takes its meet.

\paragraph{Consequence for design.} A deployment with sequence numbering but no liveness signal can
detect missing events but cannot identify unexplained silence. Conversely, a deployment with
timeout-based detection but no sequence numbering can flag streams that remain silent beyond a timeout
but cannot detect sequence gaps. Soundness guarantees must therefore specify which detection mechanisms
are available. Sect.~\ref{sec:eval} evaluates the contribution of each mechanism by removing it in turn.

\paragraph{What a beacon commits to.} A beacon from node $a$ carries a sequence high-water mark $h$ and a
time frontier $u$, the time at which it was emitted. Since nodes are authenticated and non-Byzantine and
timestamps do not decrease along a node's sequence, every event $a$ emitted with timestamp at most $u$ has
a number at most $h$, and no later event carries a timestamp at most $u$. The verifier counts the interval
up to $u$ as complete for $a$ once it holds every number up to $h$. The frontier $\tau(t)$ of
Sect.~\ref{sec:protocol} is the minimum of these $u$ over the contributing streams.

\paragraph{What the two detectors cost.} Both are implemented by the same small message. Let a
reporting policy fix, for each node and window, what that node transmits.

\begin{proposition}[Cost of accountable negative verdicts]
\label{prop:cost}
\emph{(a)} If a policy transmits nothing for node $a$ in window $w$, no sound monitor may return a
definite \noviol on a property for which $a$ could supply an otherwise unobserved witness in $w$.
Accountable negative verdicts therefore need $\Omega(1)$ messages per node per window.
\emph{(b)} Sending, once per node per window, the node's highest emitted sequence number together
with a liveness beacon preserves soundness whatever the event rate. Detection latency is at most one
window plus the detector timeout.
\end{proposition}

\begin{proof}
\emph{(a)} If nothing is sent for $a$ in $w$, the verifier's state is the same whether $a$ emitted a
witness or not. This is the indistinguishability of Theorem~\ref{thm:trilemma}, so soundness forces
\unk.
\emph{(b)} The node's highest emitted sequence number, its \emph{high-water mark}, bounds what it
emitted in the window. Any event the verifier lacks therefore shows up as a shortfall against that
number. This also covers a loss in the quiescent tail, which leaves no gap for
Lemma~\ref{lem:loss} to detect. The beacon supplies the timeout detector of
Theorem~\ref{thm:detector}: if it is absent past the timeout, the stream is marked
\textsf{unavailable}. Both detectors are live at one message per node per window, whatever the event
rate.
\end{proof}

Proposition~\ref{prop:cost} distinguishes the cost of accounting for evidence from the cost of
detecting violations. Evidence accounting requires constant reporting overhead per node per window,
whereas detecting violations requires the corresponding witness events to reach the verifier. Reducing
witness traffic limits detection capability. The verifier may then return \unk for a violation it cannot
establish, without issuing an unsupported compliance verdict.

\section{Monitoring Protocol and Soundness}
\label{sec:protocol}

Theorem~\ref{thm:detector} identifies the detection requirements for sound monitoring. We implement
these requirements in a three-tier protocol that combines sequence-based evidence accounting with
timeout-based detection of silent streams. The protocol builds on the hierarchical monitoring fabric of
our earlier work~\cite{ares2026,csr2026,critis2026}, which contributes the engineering. Unlike that
work, this paper analyses the monitoring mechanisms in terms of failure-detector requirements and
establishes their role in sound verdicts. We also introduce rules for handling membership changes
(Sect.~\ref{sec:membership}) and for stamping verdicts at the evidence frontier
(Sect.~\ref{sec:attrition}).

\paragraph{Tiers.} Each node runs a local monitor deciding the part of a specification that is local to
it, and emits events numbered contiguously in its own sequence. A \emph{group coordinator} ingests the
streams of nearby nodes, assigns a monotone ingest order and a durable identifier to each event, appends to
a durable outbox before acknowledging, and forwards upward. The \emph{verifier} consumes that output
through its own durable cursor and decides the cross-node properties. Aggregation is partial: a
coordinator resolves any property whose witness sets lie within its group and otherwise forwards the
group's projected contribution with the completeness of the streams it aggregated, so completeness
propagates with the evidence~\cite{madden2002tag}.

\paragraph{Mechanisms as detectors.} Per-node contiguous numbering plus at-least-once ingest
implements the perfect gap detector of Lemma~\ref{lem:loss}: a gap in the numbering of an ingested
stream is proof that emitted evidence is not held and marks that stream \textsf{incomplete}. The coordinator
assigns the ingest order used by the verifier. This order is compared with the emission order defined
in Sect.~\ref{sec:model}. When a discrepancy is detected, the affected verdict is marked
\textsf{degraded} to avoid evaluating the property over an incorrectly ordered
trace~\cite{lamport1978time}. The verifier-resident clock, advancing monitoring time
independently of node traffic, implements the timeout detector of Theorem~\ref{thm:detector}: a
stream that the clock shows should have reported and has not is marked \textsf{unavailable}.

\paragraph{Where the verdict is stamped.} A detector that has not yet fired leaves a window in which
a silenced stream still reads as accounted for. The protocol therefore stamps each verdict at the
\emph{evidence frontier} $\tau(t)$, the newest time to which every contributing stream is
affirmatively covered by a beacon, not at wall-clock time. Without it the protocol reports
$(\noviol,\snd)$ throughout the pre-timeout window (Thm.~\ref{thm:attrition}(ii)).

The clock must be \emph{verifier-resident}, because a liveness signal driven by the monitored nodes
shares their fate. The adversary that silences a node silences its heartbeat as well, and an empty
stream then looks like an absent one unless a signal independent of the monitored population
separates the two.

\paragraph{Isolation and backpressure.} Every event carries a globally stable identifier, so replay is
deduplicated before evaluation and a replayed witness cannot inflate a count. At the retention bound the
deployment either drops the oldest evidence silently or refuses new evidence loudly; only the second
preserves the sequence information Lemma~\ref{lem:loss} needs.

\begin{proposition}[Protocol soundness]
\label{prop:protocol}
With verdicts stamped at the evidence frontier, the protocol never commits an \sfac: every
verdict it reports as $(\noviol,\snd)$ agrees with the oracle over the interval it is stamped for.
When no event is missing it is also live: $\tau(t)$ advances and the verdict is definite after a lag
$t-\tau(t)$.
\end{proposition}

\begin{proof}[Proof sketch]
A verdict is reported as $(\noviol,\snd)$ only when every contributing stream is \snd. This requires no
unresolved sequence gap, no order violation, and no clock-flagged silence. Suppose a violation witness
was emitted but is missing from the verifier's evidence. A sequence gap detects a missing event within
an accounted-for stream, while the ordering mechanism detects events that arrive inconsistently with
the committed ingest order. If a node remains silent beyond the timeout, the verifier flags its
stream.

A further case is a node that has stopped reporting but has not yet exceeded the timeout. This is
handled by stamping the verdict at the evidence frontier $\tau(t)$. Each contributing stream must have
confirmed coverage up to that frontier through its sequence reports and beacons. An \snd composition
therefore establishes that the verifier holds all relevant events emitted within $[0,\tau(t)]$, and its
evaluation over this interval agrees with the oracle.

For liveness, suppose the contributing streams remain active, their evidence is delivered, and the
required beacons continue to arrive. The evidence frontier then advances, allowing the verifier to
issue definite verdicts for completed intervals. If a stream becomes silent, the frontier may stop
advancing, which prevents an unsupported verdict. This first appears as increasing lag and, after
the timeout, as an \unk verdict.
\end{proof}

Per event the protocol adds $O(1)$ work at each tier, and the verifier keeps $O(n)$ timers, close to the
minimum Proposition~\ref{prop:cost} requires.

\section{Membership Changes and Coordinator Failures}
\label{sec:extensions}

Changes and failures in the monitoring infrastructure can affect the reliability of runtime
verification. We examine node membership changes in Sect.~\ref{sec:membership} and coordinator
destruction in Sect.~\ref{sec:attrition}, focusing on their effects on evidence completeness, verdict
soundness, and recovery.

\subsection{Dynamic Membership}
\label{sec:membership}

The model so far fixes the node set. Real fleets do not: nodes join, depart, and are lost. This
matters for thresholds and coverage obligations, whose denominators depend on membership.

Write $M_t$ for the true membership at time $t$ and $V_t$ for the verifier's view of it. Suppose
membership is configuration rather than evidence, so $V_t$ is supplied out of band and may
diverge from $M_t$.

A \emph{reachability detector} reports which nodes can currently reach the verifier. It is not a
Chandra--Toueg crash detector, whose accuracy forbids suspecting a correct process, so it is the
observable a verifier has for a silenced but correct node.

\begin{theorem}[Membership must be reported]
\label{thm:membership}
If membership changes are not delivered on the protected path, then for threshold and coverage
properties every sound monitor abstains even in executions in which a definite verdict is warranted
and all surviving evidence is intact. This holds even when the verifier is equipped with a perfect
reachability detector.
\end{theorem}

\begin{proof}
Consider a coverage obligation requiring every sector of a set $S$ to be observed within a window,
and a node $a_i$ responsible for sector $S_j$. In execution $E_1$, $a_i$ \emph{departs} legitimately
before the window and no other node takes over $S_j$; the sector is genuinely unobserved and the
oracle verdict is \viol. In execution $E_2$, $a_i$ remains a member, observes $S_j$, and is then
silenced by the adversary before its report is delivered; the sector was genuinely observed and the
oracle verdict is \noviol.

The verifier's state is the same in both: it holds the reports of every node other than $a_i$ and
nothing from $a_i$. A perfect reachability detector does not separate them either, since in both executions
$a_i$ is permanently unreachable, which is precisely what it reports; departure and
silencing are the same event to it. The oracle verdicts differ, so any definite verdict is wrong in one of the two executions, and
soundness requires abstention. In $E_1$ the only change is the node's departure, and the coverage
question is settled. Nevertheless, the monitor must abstain because it cannot distinguish $E_1$ from
$E_2$.
\end{proof}

\begin{corollary}[Membership as evidence restores decidability]
\label{cor:membership}
If joins and leaves are delivered on the same durable, sequenced path as evidence, $E_1$ and $E_2$
become distinguishable: in $E_1$, the verifier receives a leave record carrying $a_i$'s sequence
number, so the absence of observations of $S_j$ is attributable to the departure and not to a
suppressed report, and the obligation over $S_j$, which the departure does not discharge, is decided
rather than withheld; in $E_2$, no such record exists and the missing stream is attributed by the
detectors of Sect.~\ref{sec:detectors}. Threshold and coverage verdicts then resolve exactly when
the underlying evidence is accounted for.
\end{corollary}

\begin{table}[t]
\caption{Membership under joins and departures, six missions per row (a departure and a silencing
at each of three fault times). As configuration, optimism silently clears real coverage gaps and
pessimism never resolves a question the departure settled. As evidence, neither happens.}
\label{tab:churn}
\small
\begin{tabular}{@{}lccc@{}}
\toprule
\textbf{Membership treated as} & \textbf{Silent} & \textbf{Unresolved} & \textbf{Decided} \\
 & \textbf{all-clears} & \textbf{when knowable} & \textbf{correctly} \\
\midrule
configuration, optimistic  & $24$ & $0$  & $0$ \\
configuration, pessimistic & $0$  & $27$ & $0$ \\
evidence                   & $\mathbf{0}$ & $\mathbf{0}$ & $\mathbf{24}$ \\
\bottomrule
\end{tabular}
\end{table}

\paragraph{Measurement.} We instantiate the theorem on the deployment's sector-coverage monitor, four
sectors each owned by a node, so that membership is the only possible source of unsoundness. One node
either departs, leaving its sector uncovered, or is silenced while still observing it; the verifier and
a reachability detector cannot tell the two apart, yet the verdicts differ. Table~\ref{tab:churn} shows
both failures Theorem~\ref{thm:membership} predicts. Reading persistent quiet as departure commits $24$
silent all-clears, one at every scored tick of every departure mission. Reading quiet as a present member
never issues a false clear but leaves $27$ ticks unresolved where the departure had settled the
question. With joins and leaves on the protected path both failures disappear, and genuine silencing
still yields \unk. Up to three departures per mission change nothing structurally ($11$, $8$ and
$5$ ticks); the decline reflects only the shrinking scoring window.

Membership must therefore be sequenced and durable like any other witness, since a lost leave record
recreates the ambiguity the corollary removes. A rejoining node must replay the backlog its coordinator
withheld, so that verdicts held at \unk during its absence can resolve. An integrity fault, a node that
under-reports its own activity, reduces to the loss case only if an independent corroborating witness
exists, by the argument of Theorem~\ref{thm:trilemma}.

\subsection{Coordinator Destruction}
\label{sec:attrition}

Every result so far assumes the aggregation hierarchy survives. In the motivating application it
does not: a coordinator is itself a physical node, and nodes are destroyed. We examine how
coordinator destruction affects verdict soundness, focusing on whether sequence numbers are assigned
by the nodes or by the coordinators.

\paragraph{Model extension.} Coordinators $C=\{c_1,\dots,c_m\}$ serve disjoint groups
$G_c\subseteq A$. In addition to its earlier powers, the adversary may \emph{destroy} a coordinator:
stop it permanently \emph{and} destroy the durable state it had not yet forwarded. Destruction is
strictly stronger than the crash-recovery fault studied in fault-tolerant monitoring, where the
outbox survives and is replayed on restart; a destroyed coordinator replays nothing. On destruction of
$c$, the nodes of $G_c$ re-affiliate to a successor $c'$.

\begin{theorem}[Coordinator destruction and verdict soundness]
\label{thm:attrition}
Suppose a coordinator is destroyed together with evidence it has not yet forwarded to the verifier.

\emph{(i) Coordinator-assigned numbering.} The loss of unforwarded evidence may remain unobservable.
There exist executions in which the verifier reports $(\noviol,\snd)$ despite a violation,
potentially for the remainder of the mission.

\emph{(ii) Node-assigned numbering.} The missing evidence becomes observable when the affected node's
next event or sequence-reporting beacon reaches the verifier. However, an \sfac may occur before the
loss is detected.

\emph{(iii) Verdict stamping.} A wall-clock-stamped monitor cannot soundly issue a definite verdict
for an interval whose evidence completeness is unresolved. A monitor that stamps verdicts at the
evidence frontier $\tau(t)$ instead preserves soundness by evaluating only intervals for which all
contributing streams are accounted for.
\end{theorem}

The evidence frontier $\tau(t)$ is the latest time up to which every contributing stream has confirmed
complete coverage. In fault-free operation, the verdict lag $t-\tau(t)$ is bounded by the beacon period
plus the maximum transport delay, assuming timely processing. Under sustained jamming, the frontier
may stop advancing and the lag may grow without bound. The previously established prefix verdict
remains sound for its stamped interval and has to be read as such, since it says nothing about the
execution after $\tau(t)$ and grows staler the longer the frontier stands still.

\begin{proof}
For \emph{(i)}, let the nodes of $G_c$ emit witnesses $w_1,\dots,w_k$ of a property $P$. The
coordinator ingests them, numbers them $1,\dots,k$ in its own sequence, forwards $1,\dots,j$ with
$j<k$, and is destroyed, losing $w_{j+1},\dots,w_k$; the successor begins its own sequence. The
verifier holds a contiguous prefix and a contiguous successor stream, and nothing in either
indicates the destroyed events existed, because the numbering authority that would have exposed the
gap was destroyed with them. The execution in which those events were never emitted is
indistinguishable, yet the oracles differ. Worse than in Theorem~\ref{thm:trilemma}, both streams
appear complete, so the completeness axis reports \snd and the downgrade rule does not fire: the
monitor does not abstain, it confidently clears.

For \emph{(ii)}, node-assigned numbers are not destroyed with the coordinator, so when an affected
node next reports, through the successor, the verifier sees a gap in that node's own sequence and
downgrades. But the exposure is not immediate: between the destruction and that next arrival the
verifier holds a prefix that is contiguous in every stream, so on the same reasoning as \emph{(i)}
it reports $(\noviol,\snd)$. The interval is bounded by the node's inter-report time rather than by
the mission, which is the difference from \emph{(i)}, but it is not empty.

For \emph{(iii)}, a wall-clock-stamped verdict at $t$ asserts something about the interval ending at
$t$, and by \emph{(ii)} the verifier cannot yet account for that interval's evidence, so
Theorem~\ref{thm:trilemma} applies to it directly. Stamping at $\tau(t)$ instead makes the assertion
one the verifier can support: every stream is affirmatively covered to $\tau(t)$ by a beacon, so by
the argument of Proposition~\ref{prop:protocol} the evaluation over $[0,\tau(t)]$ coincides with the
oracle's over the same interval. The monitor therefore retains a definite verdict for the latest
evidence-complete prefix, without issuing an \sfac. If the evidence frontier stops advancing, the verdict
remains sound for its stamped interval but becomes increasingly stale. Destruction stalls $\tau$ in
this way, which surfaces as growing lag and then as abstention, never as a false clear.
\end{proof}

\begin{corollary}[Distinct mechanisms for recoverable completeness and observable ordering]
\label{cor:dual}
A deployment that must recover completeness and issue definite ordered verdicts after a coordinator
is destroyed requires \emph{two} mechanisms: end-to-end node sequence numbers for completeness
accounting, and a deterministic observable order to evaluate over, which in this protocol is the
coordinator-assigned ingest order. Neither substitutes for the other. Node numbers localise missing
evidence but do not by themselves fix an order; an order assigned at the coordinator does, but
cannot survive the loss of the coordinator assigning it.
\end{corollary}

These results show that evidence completeness and event ordering require separate mechanisms.
Node-assigned sequence numbers support completeness accounting, while the coordinator-assigned ingest
order provides a deterministic ordering of events. Combining these functions in one numbering mechanism
does not provide both guarantees under coordinator destruction. The fabric this protocol builds on
assigns event identifiers at the coordinator sidecar~\cite{ares2026,csr2026,critis2026}, which by
Theorem~\ref{thm:attrition}(ii) is sound under crash-recovery but not under destruction, and switching
to the node's own number, carried end to end, costs no extra messages.

\begin{corollary}[Acknowledgement bounds what can be recovered]
\label{cor:ack}
If a node discards buffered evidence upon acknowledgement by its coordinator, destruction of that
coordinator destroys evidence that no longer exists anywhere, so the verdict can never resolve.
Recovery therefore requires that a node retain an event until it is acknowledged by the verifier, or by a tier whose survival is
independent of the node's coordinator.
\end{corollary}

\begin{table}[t]
\caption{Coordinator destruction, over six destruction times spanning the interval in which evidence
is in flight. Numbering governs how long the loss stays invisible, stamping whether a definite
verdict survives without one, and acknowledgement whether it is recovered.}
\label{tab:attrition}
\small
\begin{tabular}{@{}llccc@{}}
\toprule
\textbf{Numbering / stamping} & \textbf{Ack} & \textbf{Runs with} & \textbf{Silent} &
\textbf{Verdict} \\
 & & \textbf{an \sfac} & \textbf{window} & \textbf{resolved} \\
\midrule
coordinator, wall-clock & hop & $6/6$ & $22$ & $0/6$ \\
coordinator, wall-clock & e2e & $6/6$ & $7$  & $6/6$ \\
node, wall-clock        & hop & $6/6$ & $7$  & $0/6$ \\
node, wall-clock        & e2e & $6/6$ & $7$  & $6/6$ \\
node, frontier          & hop & $\mathbf{0/6}$ & $\mathbf{0}$ & $0/6$ \\
node, frontier          & e2e & $\mathbf{0/6}$ & $\mathbf{0}$ & $\mathbf{6/6}$ \\
\bottomrule
\end{tabular}
\end{table}

\paragraph{Measurement.} We ran the three parts of Theorem~\ref{thm:attrition} against the monitors
of the ISR deployment of Sect.~\ref{sec:eval}, so that only the transport, the destruction fault and
the stamping rule are modelled. A coordinator is destroyed mid-mission together with its unforwarded
outbox, crossing the numbering authority and the stamping rule with the acknowledgement policy
(Table~\ref{tab:attrition}). Coordinator-assigned numbering leaves the loss
invisible for the rest of the mission, a silent window of $22$ ticks; node-assigned numbering
shortens it to $7$, the affected node's inter-report time, but does not remove it, which is part
\emph{(ii)}. An earlier formulation treated node numbering as sufficient; the measurement shows it is
not. Only frontier stamping removes it, in every run, at a mean lag of $4.3$ ticks, as part
\emph{(iii)} predicts. Acknowledgement decides recovery: hop-by-hop, the destroyed evidence is gone
and the verdict never resolves; end-to-end, the surviving nodes replay and every run resolves.

\paragraph{Over live brokers.} We repeated the three configurations over Mosquitto and NATS JetStream,
killing the coordinator process with \texttt{SIGKILL} together with its unforwarded outbox. Frontier
stamping again commits no silent all-clear in any run; the wall-clock configurations vary ($4$ to $6$ of
$6$ runs, a window of $3$ to $6$ ticks), since live delivery timing decides how long the loss stays hidden.

A monitor that always abstained would be sound but never definite; frontier stamping adds a definite
verdict over the covered prefix, at a bounded lag.

\paragraph{What a successor may conclude.} A coordinator $c'$ taking over at time $t$ knows neither
what its predecessor ingested nor what it forwarded. It publishes a \emph{handover-gap marker}
covering the interval from its predecessor's last verifier-acknowledged position to $t$, and the
verifier treats every stream in $G_c$ as \textsf{incomplete} over that interval until the nodes' own
numbering closes it. Verdicts spanning the interval are held at \unk and resolve as the surviving
nodes replay from their last end-to-end acknowledgement (Cor.~\ref{cor:ack}), so the mission verdict
shows a gap where the destruction happened instead of an unsupported compliance verdict.

\section{Experimental Evaluation}
\label{sec:eval}

\paragraph{Setup.} The measurements come from a deployment monitoring an LLM-planned ISR swarm of
four platforms over live brokers, MQTT (Mosquitto) on the node-to-coordinator hop and NATS JetStream
on the durable hop to the verifier, with the node tier exercised over ROS\,2 and ArduPilot
SITL~\cite{ares2026,csr2026,critis2026,mesas2026swarm,mqtt,nats}. The monitored obligations are three
cross-node properties of that mission: a prohibited collective intelligence package, a swarm-wide emission budget,
and entry to a sensitive zone only after authorisation. The fault campaign is fault-free, a dropped
witness and a jammed platform, giving nine oracle incidents. The monitors, drivers and finite-model
checker will be released on acceptance.

Five experiments are reported: two reused from that deployment (the baseline comparison and the
guarantee ablation) and three new (the membership and destruction sweeps of Tables~\ref{tab:churn}
and~\ref{tab:attrition}, and the bandwidth sweep below). The new experiments run against the
deployment's own monitor code, with only the transport, the fault and the design choice under test modelled.

\paragraph{The trilemma, and its two failure classes.} Over the campaign, per-node guardrails commit
$9$ silent false all-clears out of $9$ oracle incidents, a best-effort central monitor $3$, and the
protocol of Sect.~\ref{sec:protocol} none. The guardrail figure is Theorem~\ref{thm:trilemma}'s locality
escape in reverse: every witness set spans nodes, so no local monitor holds one and each node is
compliant while the system is not. The central monitor's three are the trilemma proper: it consumes
the merged stream but cannot report that part of it is missing, and a two-valued domain leaves only
the optimistic answer.

Withdrawing the two detectors one at a time separates the failure classes. Without sequence-gap
detection the protocol commits one silent all-clear under the dropped witness and none under the jammed
node; without the verifier-resident clock it commits none under the dropped witness and two under the
jammed node, so each mechanism covers one failure class and neither covers both.

\begin{table}[t]
\caption{The cost of accountable negative verdicts, at $4$ events per node per window over the fault-free,
dropped-witness and jammed-node runs, over the scored ticks of all three. Shedding witnesses costs
detection; shedding the $\Theta(1)$ summary permits silent false all-clears and saves little.}
\label{tab:bw}
\small
\setlength{\tabcolsep}{2.5pt}
\begin{tabular}{@{}lrccc@{}}
\toprule
\textbf{Reporting policy} & \textbf{B/node/} & \textbf{Silent} & \textbf{Abstained} &
\textbf{Detect} \\
 & \textbf{window} & \textbf{all-clears} & & \textbf{lag} \\
\midrule
all witnesses \& summary    & $364$ & $\mathbf{0/69}$ & $26/69$ & $1$ \\
1-in-2 witnesses \& summary & $192$ & $\mathbf{0/69}$ & $69/69$ & never \\
1-in-4 witnesses \& summary & $108$ & $\mathbf{0/69}$ & $69/69$ & never \\
summary only                 & $24$  & $\mathbf{0/69}$ & $69/69$ & never \\
all witnesses, no summary    & $340$ & $25/72$ & $0/72$ & $1$ \\
dark (nothing sent)          & $0$   & $36/72$ & $0/72$ & never \\
\bottomrule
\end{tabular}
\end{table}

\paragraph{The cost of accountable verdicts.} Table~\ref{tab:bw} sweeps the reporting policy against the
deployed package monitor, with verdicts stamped at the evidence frontier as
Theorem~\ref{thm:attrition}(iii) requires, so that the policy is the only variable. An event report
is charged $96$\,B and a summary message, carrying the node's high-water sequence number and its
liveness beacon, $24$\,B.

The two costs separate cleanly. Every policy retaining the summary commits no silent all-clear at
any bandwidth, down to the $24$\,B per node per window that is the summary alone. What it loses as
witness traffic is shed is detection: verdicts move to \unk, from $26$ of $69$ scored ticks at full
witness traffic to all $69$ once decimation removes a part of the package from the shipped stream.
The policy that economises the other way keeps every witness and drops the summary, saving $24$ of
$364$\,B per node per window, under seven per cent, and commits $25$ silent all-clears. Both faults
are of the kind only the summary catches: the dropped witness is the affected node's last event, so
no subsequent arrival can expose the gap and only the high-water number can, and the jammed node has
no beacon whose absence could be noticed. The dark policy transmits nothing and clears silently at
every tick after the package completes, which is Proposition~\ref{prop:cost}'s $\Omega(1)$ bound
made concrete.

Sweeping the event rate from $1$ to $16$ events per node per window leaves the summary at $24$\,B while
witness traffic grows from $100$ to $1312$\,B, so its share falls from $19\%$ to under $2\%$. The
policies retaining the summary commit no silent all-clear at any rate; the policy dropping it commits $25$
at every rate.

\paragraph{What the protocol costs.} Over the live brokers the verdict path has a $3.7$\,ms median
over $n{=}1000$ events, and a $42$\,ms median under a $10\%$-drop link with durable retransmit and no
accepted evidence lost. Mission-monitor evaluation grows roughly linearly and stays under $2$\,ms to
$1000$ nodes. Frontier stamping adds no latency but a verdict lag, $4.3$ ticks on average in the
destruction sweep, close to the beacon period plus transport delay.

\paragraph{Threats to validity.} The three new sweeps model the transport but call the deployed monitors, so
property semantics, completeness accounting and the downgrade rule are those of the running system;
timing is simulated and the byte model nominal. The swarm has four platforms and the scale result is a synthetic trace. The core model
assumes authenticated non-Byzantine nodes (Sect.~\ref{sec:membership}) and monotone per-node timestamps, which the
experiments script and do not enforce. The oracle essentially executes the same property code on the pre-transport trace. This is why the
comparison tests the evidence channel and not the specification.

\section{Related Work}
\label{sec:related}

\paragraph{Distributed Runtime Verification} Distributed runtime verification has been studied
under several monitoring architectures and communication
assumptions~\cite{bauer2012decentralised,elhokayem2017decentralized,francalanza2018distributed}.
Some approaches distribute the decision across nodes and tolerate monitors that crash or hold
different opinions about the execution~\cite{bonakdarpour2016decentralized,falcone2021distributed,fraigniaud2014opinions}.
Spatio-temporal properties have also been
monitored~\cite{strel2017}. The most closely related work addresses incomplete observations, network
failures, and out-of-order delivery~\cite{basin2015failure,basin2017outoforder,joshi2017lossy,stoller2011estimation}.
It provides semantics and algorithms for evaluating a property when observations are unreliable. Our
work addresses an earlier step. Before a definite verdict is issued, the verifier must establish that its
evidence is complete. We study how sequence reports, timing information, and membership records support
this decision when an adversary suppresses messages selectively. Policy auditing over incomplete
logs~\cite{garg2011audit} addresses a related problem, but in a different setting and with a different
treatment of evidence availability.

\paragraph{Failure Detectors} Chandra and Toueg classified failure detectors by accuracy and
completeness~\cite{chandra1996unreliable} and identified the weakest detector for
consensus~\cite{chandra1996weakest}. Process failure is a different problem from ours. A node whose link
to the verifier is blocked keeps running and keeps emitting events, so knowing that the node is alive says
little about whether its evidence has arrived. The detector we need reports whether a node has emitted
events that the verifier does not hold. Theorem~\ref{thm:detector} states what such an
evidence-completeness detector must provide for sound monitoring, without placing it in the
Chandra--Toueg hierarchy.

\paragraph{Impossibility Results and Multi-Valued Verdicts} Indistinguishability arguments underlie
many impossibility results in distributed computing, FLP among them~\cite{fischer1985impossibility}. Ours
concerns a different situation, a single verifier that must answer by a deadline while relevant events may
have been suppressed. Multi-valued verdicts and monitorability are well studied in runtime
verification~\cite{bauer2011runtime,falcone2012monitorability,leucker2009brief}, where an inconclusive
verdict means that the trace seen so far does not yet settle the property. We add a second reason for
abstaining, namely that the trace seen so far may not contain every relevant event.
Fraigniaud et al.~\cite{fraigniaud2014opinions} prove that fault-tolerant decentralised monitoring needs
three opinions because monitors may disagree, whereas the three values here arise from suppression at
a deadline-bounded verifier.

\paragraph{Monitoring Tools and Applications} MonPoly~\cite{monpoly} and the stream-based
engines~\cite{dejavu,rtlola,tessla} evaluate temporal specifications over event streams. We address a
separate concern, whether the evidence delivered to such engines is complete. The hierarchical forwarding in our
protocol resembles in-network aggregation as in TAG~\cite{madden2002tag}, with the addition that each
forwarded summary carries the completeness of the streams behind it. Runtime monitoring for
robots~\cite{ferrando2021rosmonitoring} and work on autonomous swarms~\cite{chung2018aerial} set the
application context, and the adversarial scenarios draw on security concerns about LLM-based systems and
the composition of their actions~\cite{agentdojo2024,greshake2023injection,cpsiotsec2026}.

\section{Conclusion}
\label{sec:conclusion}

Distributed runtime verification can stay sound when its evidence is incomplete or deliberately
suppressed, under conditions this paper has set out. For the class of cross-node properties we study,
no asynchronous monitor is sound, decisive, and available at once, so a monitor that keeps soundness
has to abstain whenever the evidence does not support a definite verdict.

Accounting for missing evidence takes several mechanisms, and they are not interchangeable. Sequence
numbering and periodic reports expose missing events. Silent streams can only be detected under timing
assumptions. Membership changes have to be reported through the protected evidence channel, and
coordinator destruction adds its own requirements on event numbering, verdict stamping, and
acknowledgement scope.

The experiments show what the guarantees cost. Constant-size sequence summaries keep evidence accounting
independent of the event rate. Cutting witness traffic can stop a violation from being detected, and the
sound monitor then returns an inconclusive verdict in place of unsupported compliance. Stamping verdicts
at the evidence frontier keeps them sound at the expense of a lag behind the current execution.

All of this rests on authenticated, non-Byzantine nodes and on the ordering and timestamp assumptions of
the protocol. Relating evidence-completeness detection to the Chandra--Toueg failure-detector hierarchy
remains open.

\balance
\begin{acks}
This work has received funding from the European Union's Digital Europe Programme (CYBERGUARD, No.~101190251).

The authors used Claude (Anthropic) to support manuscript review and experiment automation. The
authors checked all content and take full responsibility for it.
\end{acks}

\bibliographystyle{ACM-Reference-Format}
\bibliography{references}

\end{document}